\documentclass[a4paper,12pt]{article}
\usepackage{graphicx}

\begin{document}

\begin{center}
{\large \bf The Sensitivity of a Lithium Experiment\\ on Solar
Neutrinos to the Mixing Angle $\theta _{12}$.}

\vskip 0.2in

Anatoly Kopylov and Valery Petukhov\\ Institute of Nuclear
Research of Russian Academy of Sciences \\ 117312 Moscow, Prospect
of 60th Anniversary of October Revolution 7A
\end{center}

\begin{abstract}
A lithium-based radiochemical detector is aimed primarily to
detect neutrinos from CNO cycle what will provide a direct proof
of its existence and will be a stringent test of the theory of
stellar evolution. Another task which can be solved by this
experiment is to measure a mixing angle $\theta _{12}$. The
sensitivity of a lithium experiment to $\theta _{12}$ was
calculated by Monte-Carlo following the proposed original
technique which can be used as a complimentary one to a chi-square
technique usually applied to this task. It is shown that the
accuracy of measurement of the mixing angle in solar neutrino
experiments is principally limited by the accuracy of a lithium
experiment due to the limiting uncertainty of the energy generated
in a pp chain.
\end{abstract}

This paper describes the procedure to calculate the accuracy in
the evaluation of the mixing angle $\theta$ (in this paper we are
considering only a mixing angle $\theta _{12}$) if the capture
rate of solar neutrinos is measured by a lithium detector with
certain accuracy. The general idea of this procedure was proposed
in \cite{1}, here the computer implementation of this idea and the
conclusions drawn on a basis of the results obtained are
presented. First part describes the procedure. The interested
reader may skip this part and start reading part 2 which presents
the results obtained and their discussions. The conclusions
important for the practical questions of the realization of a
lithium experiment are formulated in part 3. \vskip 0.2in

\textbf{1. The procedure to calculate}. \vskip 0.1in

The main point of this technique is that a relatively modest
accuracy ($\sim $10\%) in the measurement of the neutrino capture
rate in a lithium experiment will evaluate with a very high
accuracy ($\sim $0.5\%) the energy generated in a hydrogen chain
what combined with the luminosity constraint proposed by M.Spiro
and D.Vignaud \cite{2} will determine precisely the mixing angle
$\theta$. And vice versa: if the energy generated in a CNO cycle
is not measured this will be principal limiting factor in the
evaluation of $\theta $. Here we take $\Delta m^2 = 7.3 \times
10^{-5}$ $eV^2$ of the best fit point and take it as an accurately
known value taking into consideration that very soon it will be
measured with a very good accuracy in experiment KamLAND. The
procedure begins with the simulation of two values: the true
capture rate in a lithium experiment for a given measured capture
rate and $tan^2 \theta$ within the interval limited by the global
rates of solar neutrino experiments \cite{3} and KamLAND \cite{4}
as it was shown in a number of papers \cite{5} and most recently
in \cite{6}. It was taken here:

\begin{center}
$tan^2\theta  = 0.42 \pm 0.08$ $(1 \sigma)$
\end{center}

\noindent and the Gaussian distribution by the simulation of
random $\theta _i$. It was taken also that the best fit point of
$\theta$ corresponds to the fluxes of solar neutrinos given by a
standard solar model of BP2000. We should stress at this moment
that what are exactly the values taken on the input of the
procedure is not very critical for the final result, as it will be
shown later. As a next step it was necessary to find the
contribution of neutrinos generated in a CNO cycle to the total
rate found in experiment with a lithium target. It is a
straightforward procedure of subtracting the effect for the known
fluxes of neutrinos generated in a pp-chain and for the calculated
factor for the electron neutrinos to survive. This factor was
calculated for two different oscillation scenarios for two
different energy regions: vacuum oscillations with the factor
$P_{ee} = 1 - 0.5 sin^2 2 \theta$ and MSW oscillations with the
factor $P_{ee} = tan^2 \theta$. The boarder between these two
regions was taken to be 2 MeV, this is an arbitrary supposition
taken just for convenience. This supposition does not contradict
to the experimental data so far, it will be corrected by the
future experiments, but for the aim of the present paper this
choice has no principal meaning. Let's just keep in mind that by
the time this question is cleared the results of the calculations
should be corrected. The rate was calculated for each neutrino
source of pp chain: $^7$Be, pep and $^8$B neutrinos according to
the formula

\begin{center}
$U_a(\theta) = f_a \smallint dE \cdot P_{ee}(\theta,E)\cdot \Phi
_a^{BP}(E)S_{Li}(E)$ \hskip 0.5in (1)
\end{center}

Here $a$ denotes $^7$Be, pep and $^8$B neutrinos, $f_a$ is a
reduced (relative to the one given by BP2000 model) neutrino flux
which is taken 1.0 at the beginning and later on we will see the
effect if it is fluctuated around 1.0 with the uncertainty $\sigma
$; $P_{ee}(\theta,E)$ is the probability for the electron neutrino
to survive; $\Phi _a^{BP}(E)$ are the neutrino fluxes given by
BP2000; $S_{Li}(E)$ is the cross section of neutrino capture on
lithium calculated by J.Bahcall \cite{7}. After subtracting the
effect from neutrinos of a pp chain the rate from CNO neutrinos
was obtained. The ratio of the effect found for a CNO cycle to the
calculated one by the formula (1), where $a$ means $^{13}$N or
$^{15}$O neutrinos, will be denoted further by a factor G. If we
take the ratio of $^{13}$N to $^{15}$O neutrinos generated in a
CNO cycle fixed and equal to what BP2000 suggests, then the factor
G means that the fluxes of these neutrinos generated in the Sun
are equal to the ones given by BP2000 multiplied by the factor G.
But then, to fulfill the luminosity constraint, one should
introduce another factor D for the fluxes of neutrinos generated
in a pp chain according to the equation of the luminosity balance

\begin{center}
0.015 G + 0.985 D =1 \hskip 0.5in (2)
\end{center}

\noindent for CNO cycle and for a pp chain and

\begin{center}
$0.913f_{pp} + 0.002f_{pep} + 0.07f_{Be} + 0.0071f_N + 0.0079f_O =
1$ \hskip 0.5in (3)
\end{center}

\noindent for all neutrino sources with the coefficient by the
reduced neutrino flux greater 0.0001. The coefficients by$f_{pp}$,
$f_{pep}$ and $f_{Be}$ were obtained from numbers of Table 1
presented in \cite{8}, the coefficients by $f_N$ and $f_O$ were
calculated accounting that the energy produced for each $^{13}$N
neutrino in a first half-cycle CNO

\begin{center}
$\alpha(^{13}N) = M(^{12}C) + 2M(^1H) - M(^{14}N) - \langle
E_{\nu} \rangle (^{13}N) = 11.00$ $MeV$
\end{center}

\noindent and for each $^{15}$O neutrino for the second half-cycle
CNO

\begin{center}
$\alpha(^{15}O) = M(^{14}N) + 2M(^1H) - M(^4He) - M(^{12}C) -
\langle E_{\nu} \rangle (^{15}O) = 14.01 MeV$
\end{center}

From the expression (2) it follows that the energy generated in a
CNO cycle is 1.5\% times G of the total energy generated in the
Sun. We take here a fixed ratio of He3-He4 termination chain
relative to $^3$He - $^3$He termination chain suggested by BP2000
model because the flux of $^7$Be neutrinos has not been measured
by the present time with the accuracy sufficient to establish
unambiguously this ratio. For the purpose of this paper it is not
very essential moment. So from this expression we find the factor
D as a one by which the luminosity constraint is fulfilled. Lets
note that the factor D is just the ratio of the real energy
generated in a pp chain to the one suggested by a BP2000 model.
And while the factor G can be large, for example, it can be 1.5 or
nearby, the factor D differs from 1.0 only by a small quantity of
the order of 1\% . Because the fluctuations in $tan^2 \theta$ are
taken rather large ($\sigma $ = 20\% ) sometimes the value G
becomes negative what obviously has no physical meaning. These
events will be rejected so that on the output the number of the
real simulations is indicated out of 1000 tries. This is not very
good approach from the statistical point of view, but for our
study it has no dramatic consequences. It reminds the old story
about Runs with negative numbers (after subtracting the
background) in Davis experiment. And now we are coming to the
final point. For each value of D found in each simulation we find
the new $tan^2 \theta^{\prime}$ which agrees with the luminosity
constraint applied only to a pp chain. Here, at first, we don't
modify the relative structure of the neutrino sources inside the
chain, we take it according to BP2000 but the absolute values of
neutrino fluxes in the zone of their generation become higher or
lower by the factor corresponding to a new $\theta^{\prime} $ and
to the energy of the neutrino source. Then we let the fluxes of pp
neutrinos (here we take the flux of pp neutrinos together with the
flux of pep neutrinos using the fact that the ratio of pep
neutrinos to pp neutrinos is well known) or $^7$Be neutrinos, or
$^8$B neutrinos, or the combination of them, to vary with a
certain $\sigma $ and see what will be the effect. \vskip 0.2in

\textbf{2. The results obtained and their discussion.} \vskip
0.1in

Fig.1 presents the summary of the calculations.

\begin{figure}[h!]
\centering
\includegraphics[width=4in]{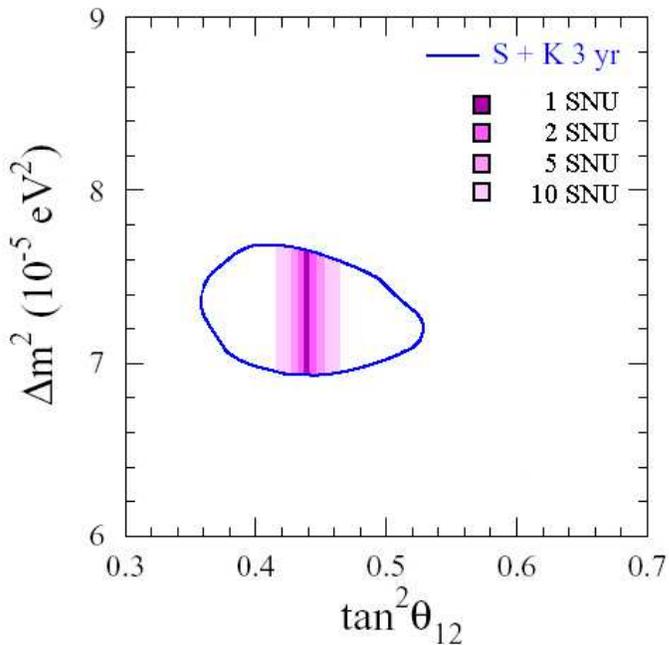}
\caption{The sensitivity plot of a lithium detector to
$\theta_{12}$ for different experimental uncertainties $\sigma$.
The contour S+K 3 yr was taken from \cite{6}}
\end{figure}

\noindent One can see that the limit on $\theta^{\prime}$ depends
critically upon the accuracy obtained in a lithium experiment. The
data presented on this figure were obtained for the extreme case
when the fluxes of neutrinos from a hydrogen chain are exactly
known, i.e. the uncertainty in their evaluation is zero, and they
are equal to the ones suggested by BP2000 model. How this result
was obtained one can see on Fig.2 which shows the data for
different accuracies of the rate measured in a lithium experiment
(1$\sigma $ = 1 SNU, 2 SNU, 5SNU and 10 SNU) for the average rate
21.7 SNU expected in experiment for the oscillation parameters
listed above.

\begin{figure}[h!]
\centering
\includegraphics[width=2.5in]{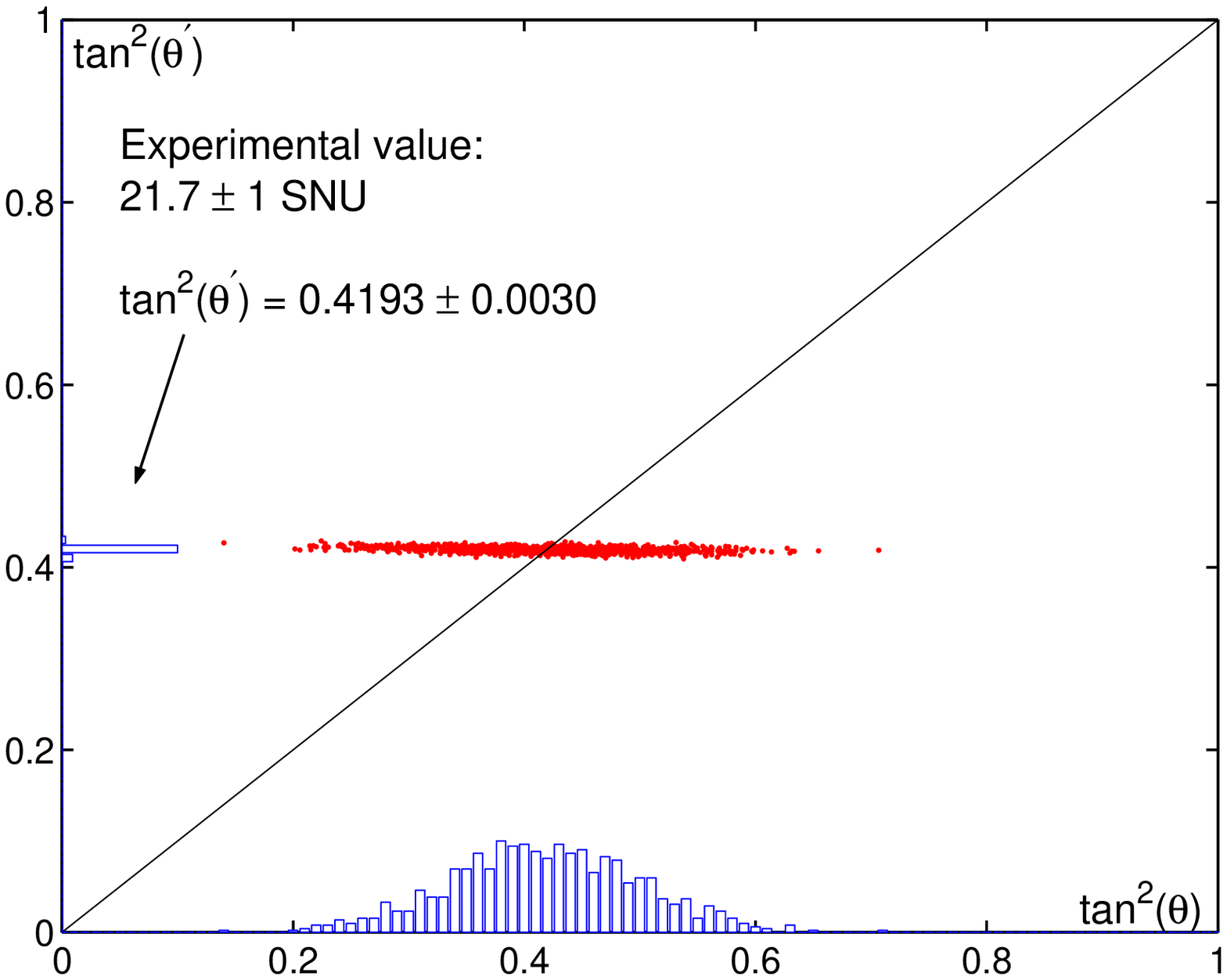}
\includegraphics[width=2.5in]{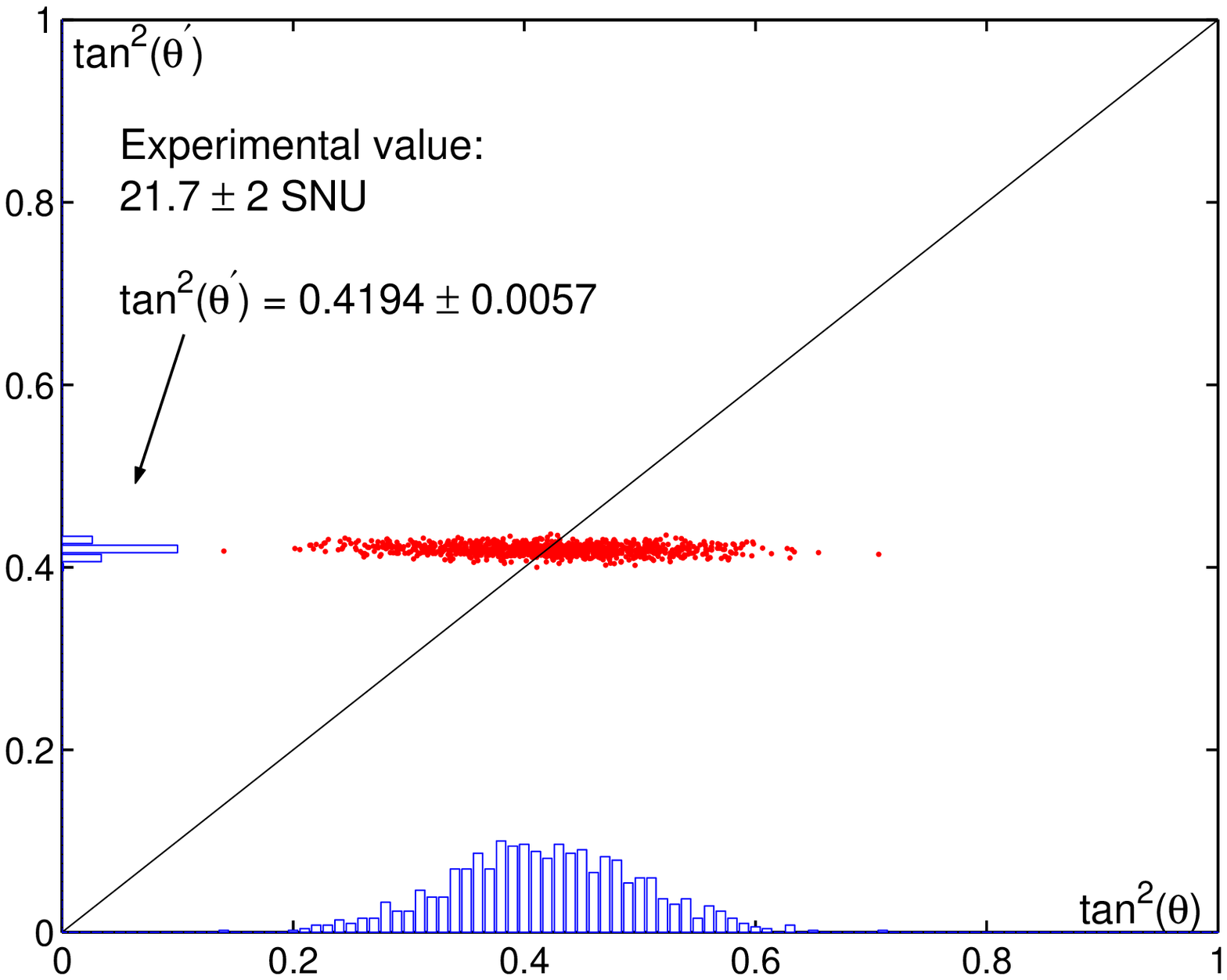}
\includegraphics[width=2.5in]{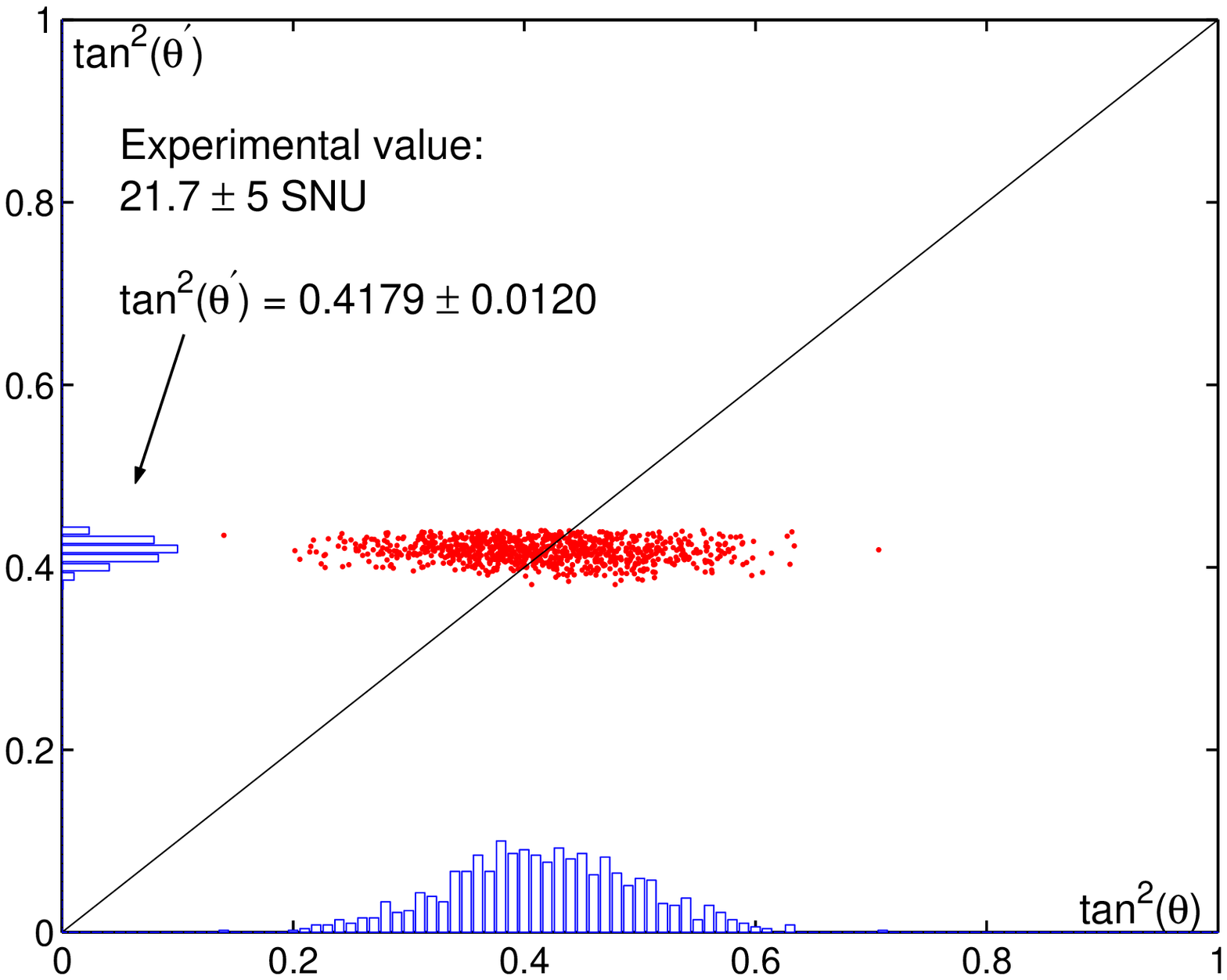}
\includegraphics[width=2.5in]{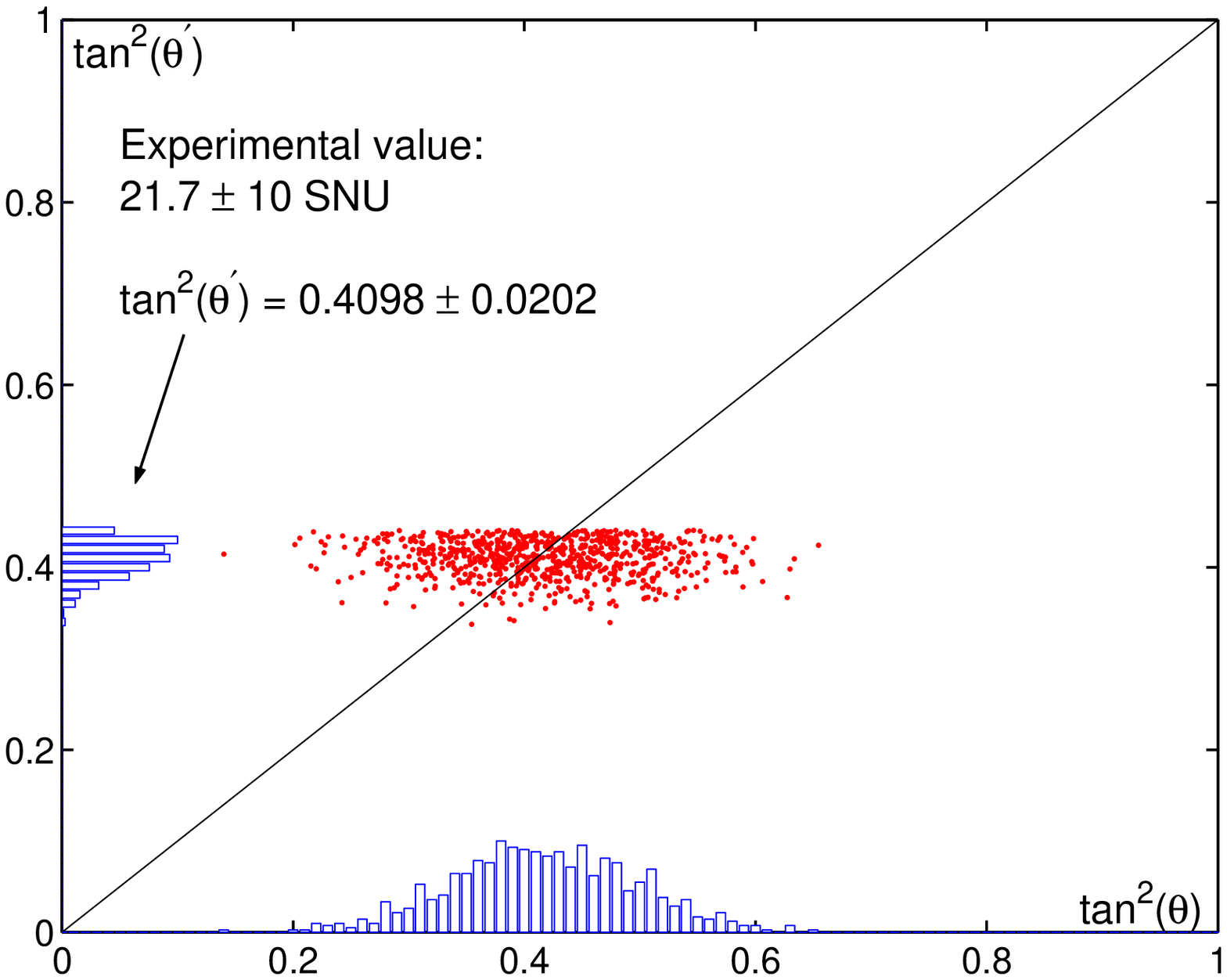}
\caption{The accuracy of the evaluation of $\theta^{\prime}$ for
different uncertainties of a lithium experiment. The distribution
on Y-axis is obtained with the Luminosity Constraint applied to
the data.}
\end{figure}

\noindent Two values were simulated: the neutrino capture rate
measured by experiment and a mixing angle $\theta $. The value of
$tan^2 \theta^{\prime}$ is found as a one to comply with the
luminosity constraint. One can see the scattering of the simulated
points and histograms for $tan^2 \theta$ and $tan^2
\theta^{\prime}$ distributions. The most interesting result which
is clearly presented on all figures is that the distribution for
$tan^2 \theta^{\prime}$ is much narrower than for $tan^2 \theta$.
This is a clear demonstration of the power of the luminosity
constraint if the fluxes of CNO neutrinos are measured. The data
show that the higher is the accuracy of a lithium experiment the
more limiting is the result for the energy generation in a pp
chain and the more precisely is determined the mixing angle
$\theta^{\prime}$. For large experimental uncertainty 10 SNU the
points get scattered in a large field and the distribution on
$tan^2 \theta^{\prime}$ is asymmetrical. As one can see later this
case of 10 SNU is an extreme one, in the real experiment one can
expect as very realistic the accuracy of 5 SNU which can be
achieved in the time scale of 1 year even with the very simplified
counting system. The accuracy of 1 SNU can be achieved with the
counting system on a basis of a cryogenic detector which would
enable to achieve the efficiency of counting of $^7$Be close to
100\% \cite{10}.

\begin{figure}[h!]
\centering
\includegraphics[width=2.5in]{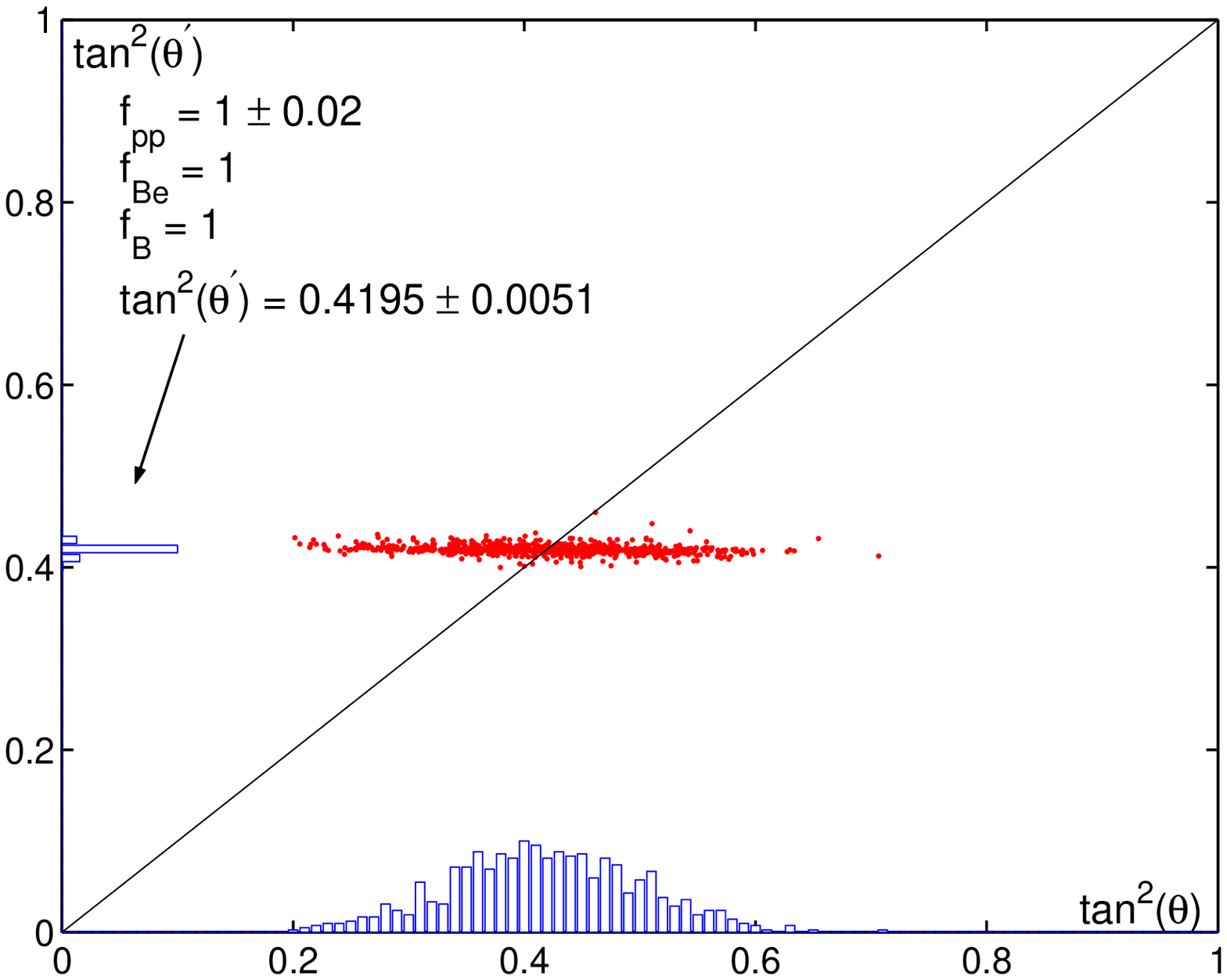}
\includegraphics[width=2.5in]{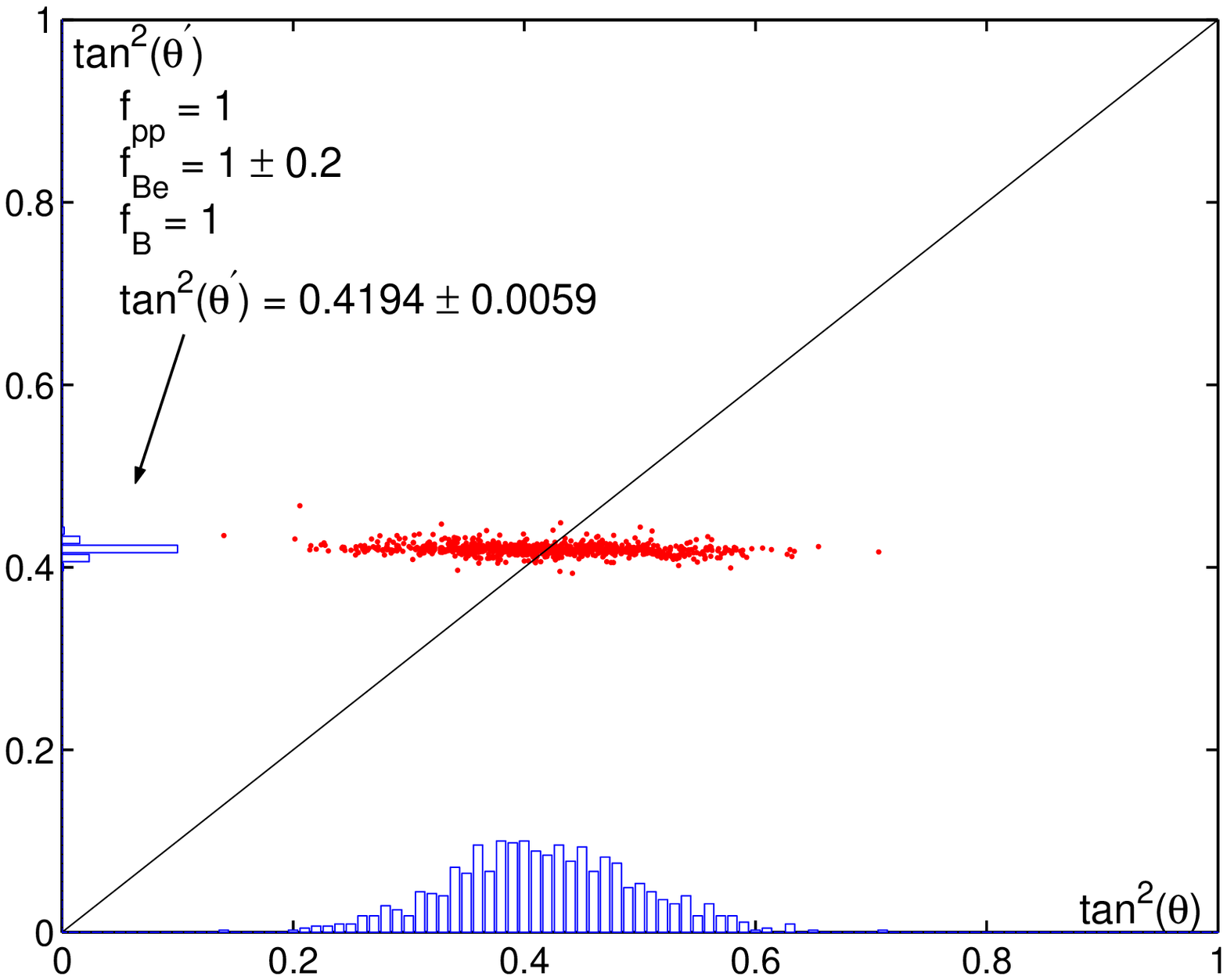}
\includegraphics[width=2.5in]{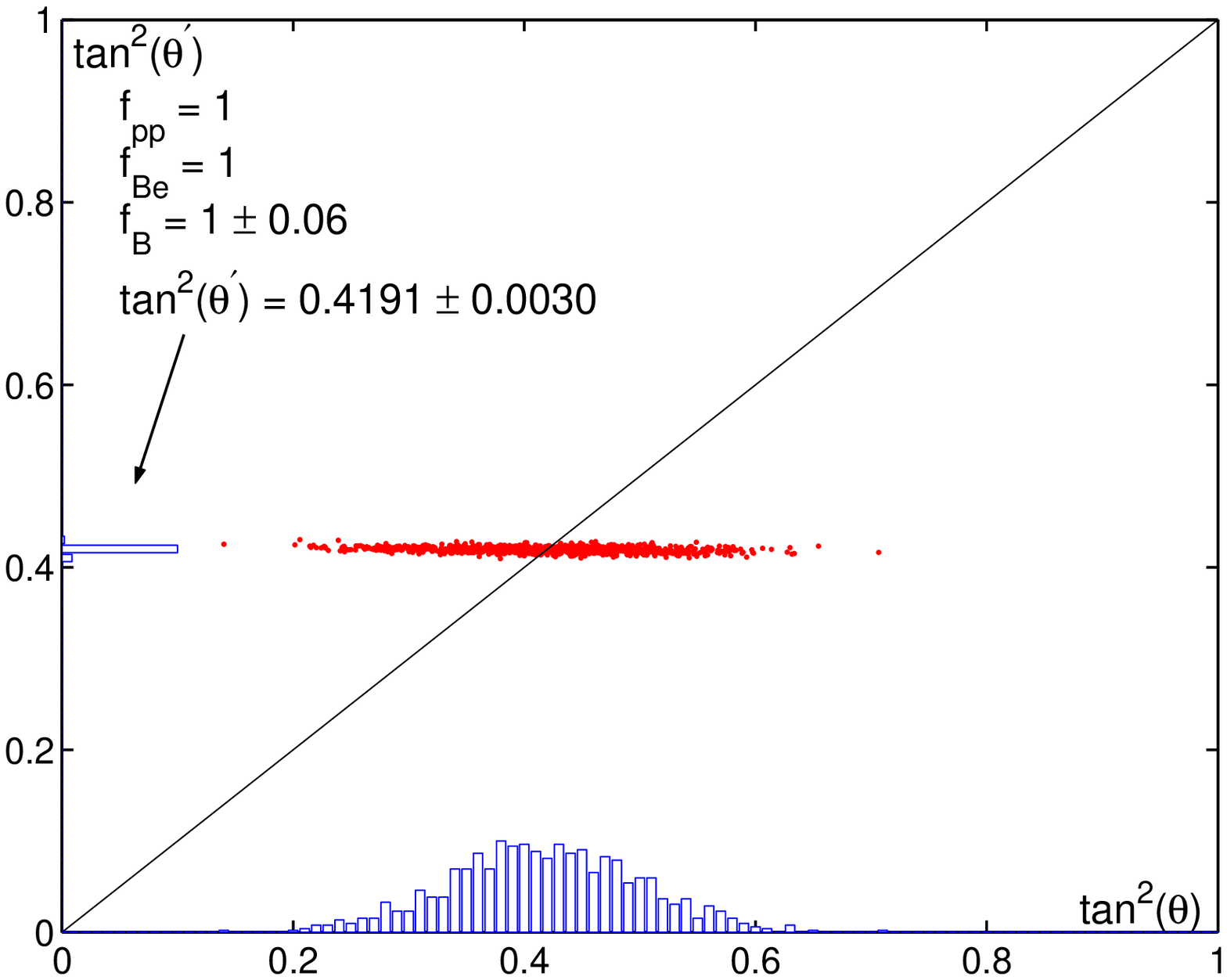}
\includegraphics[width=2.5in]{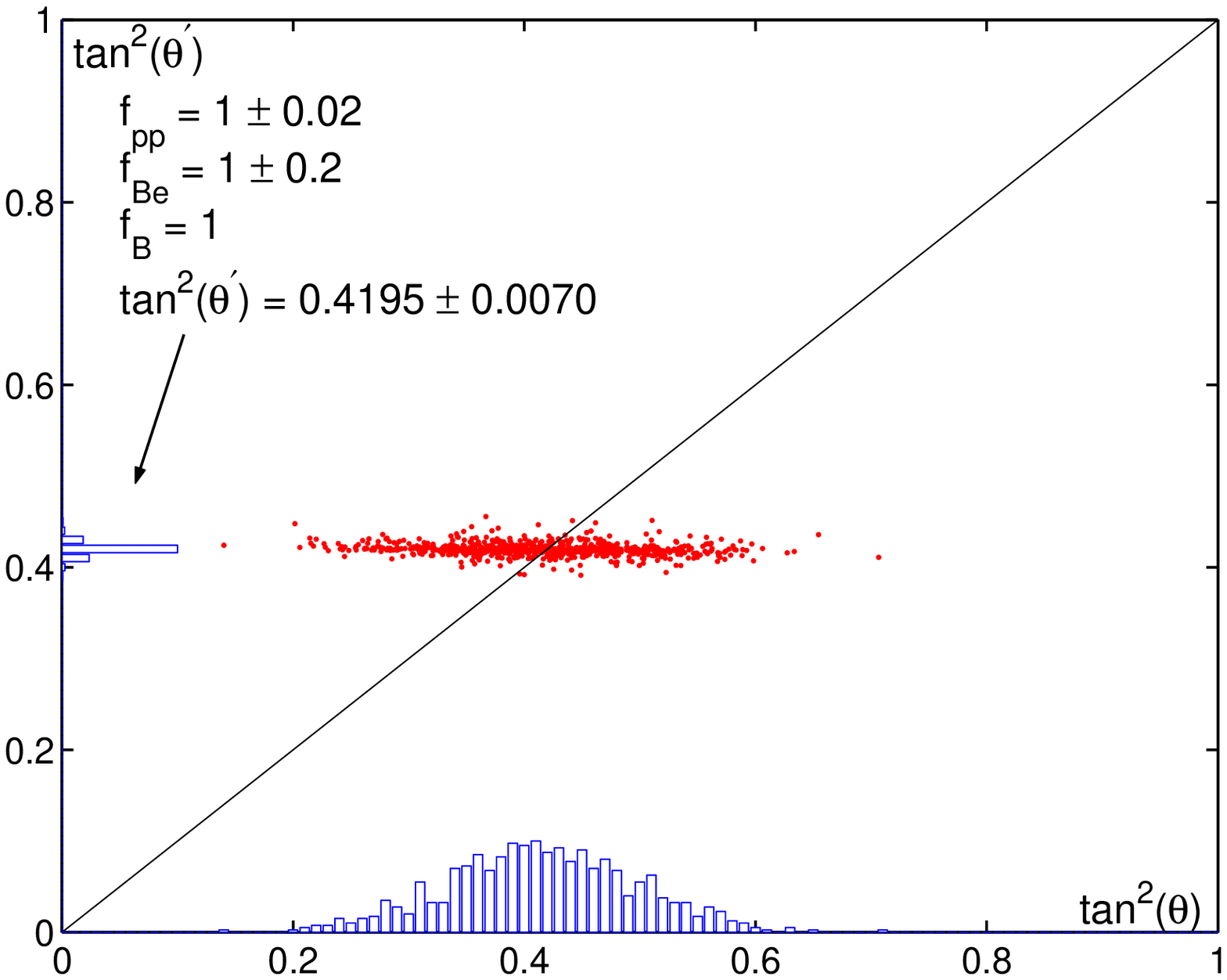}
\caption{The same as Fig.2 but the flux of neutrinos of pp-chain
are varied.}
\end{figure}

    Fig.3 shows the data similar to Fig.2 only here the flux of pp
neutrinos is varied with $\sigma $ = 2\% , the flux of $^{{\rm
7}}$Be neutrinos is varied with $\sigma $ = 20\% , the flux of
$^8$B neutrinos is varied with $\sigma $ = 6\% , both pp neutrinos
and $^7$Be neutrinos are varied with $\sigma $ = 2\% (pp) and with
$\sigma $ = 20\% ($^7$Be) for the same average rate 21.6 SNU.
Comparing the data presented on these figures with the ones
presented on Fig.2 one can see how the uncertainties of the
different neutrino fluxes change the result for $tan^2
\theta^{\prime}$. The influence of boron neutrinos is not so
strong, the influence of pp and $^7$Be neutrinos is quite
substantial. Next two figures are similar to the Fig.2 only the
data were obtained not for the average expected rate in a lithium
experiment 21.7 SNU, but for 21.7 $\pm $ 5 SNU, Fig.4 for 16.7 SNU
and Fig.5 for 26.7 SNU. One can see that the general picture is
not changed drastically for these cases. Figure 6 shows how $tan^2
\theta^{\prime}$ depends on the neutrino capture rate R measured
in a lithium experiment.

\begin{figure}[h!]
\centering
\includegraphics[width=2.5in]{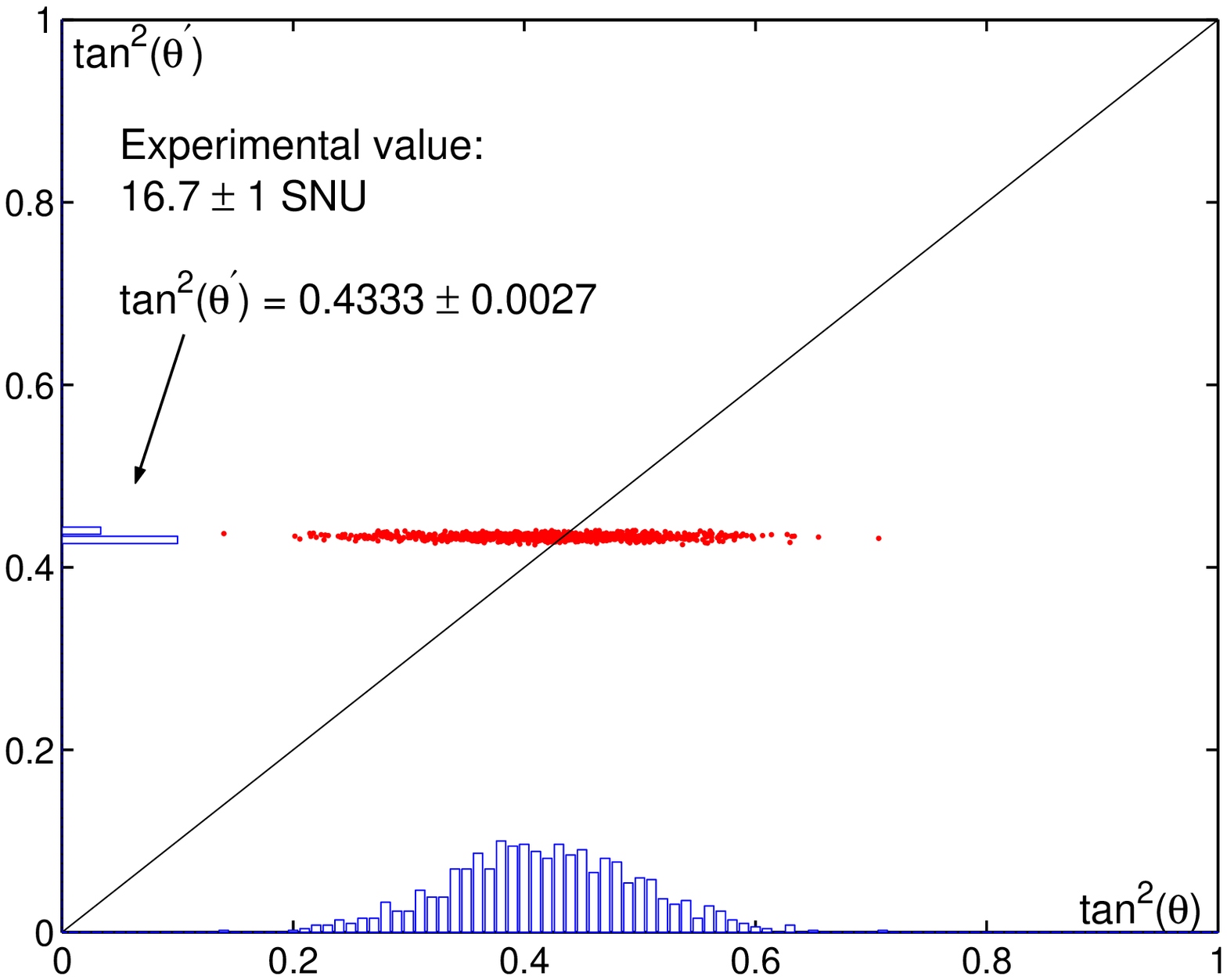}
\includegraphics[width=2.5in]{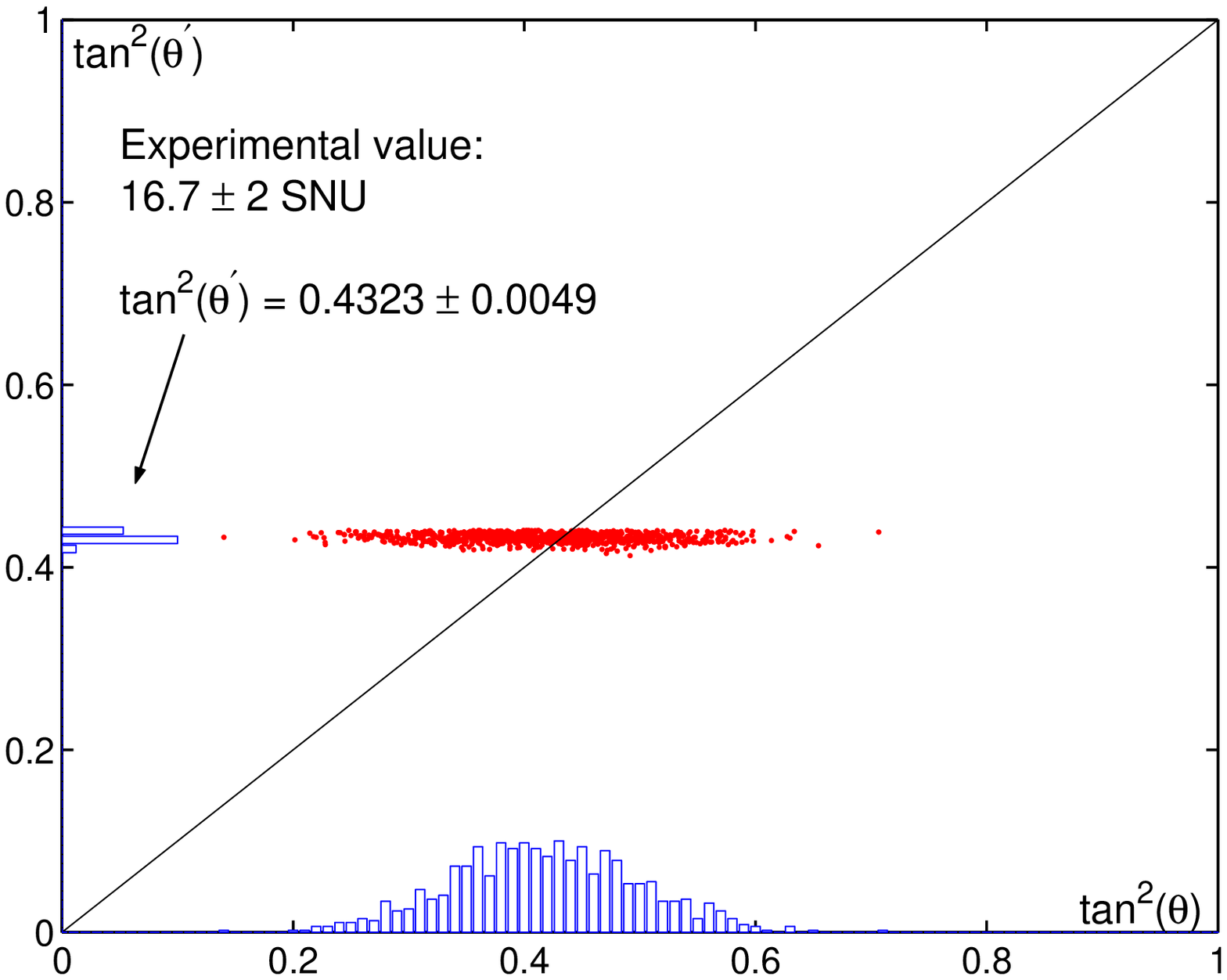}
\includegraphics[width=2.5in]{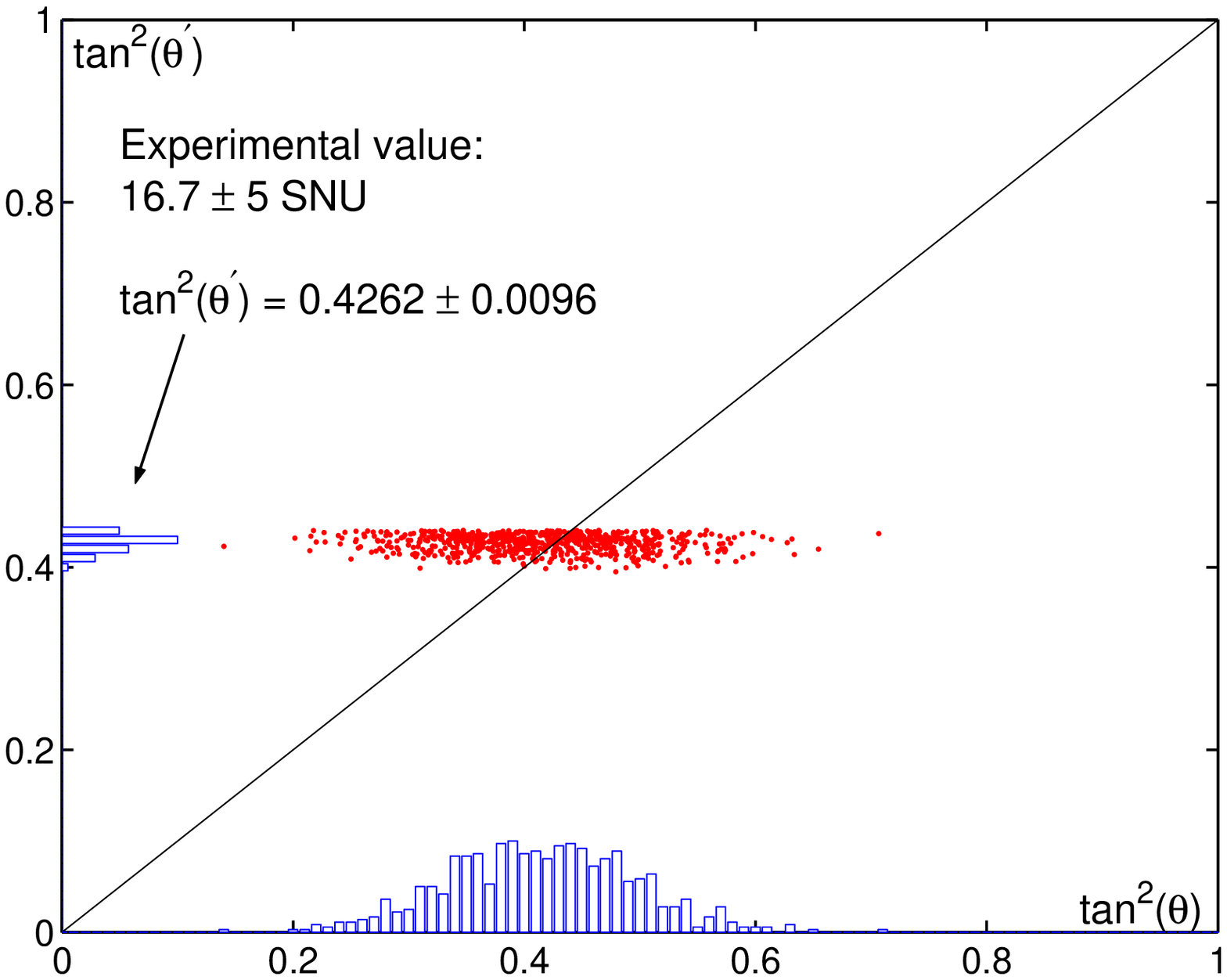}
\includegraphics[width=2.5in]{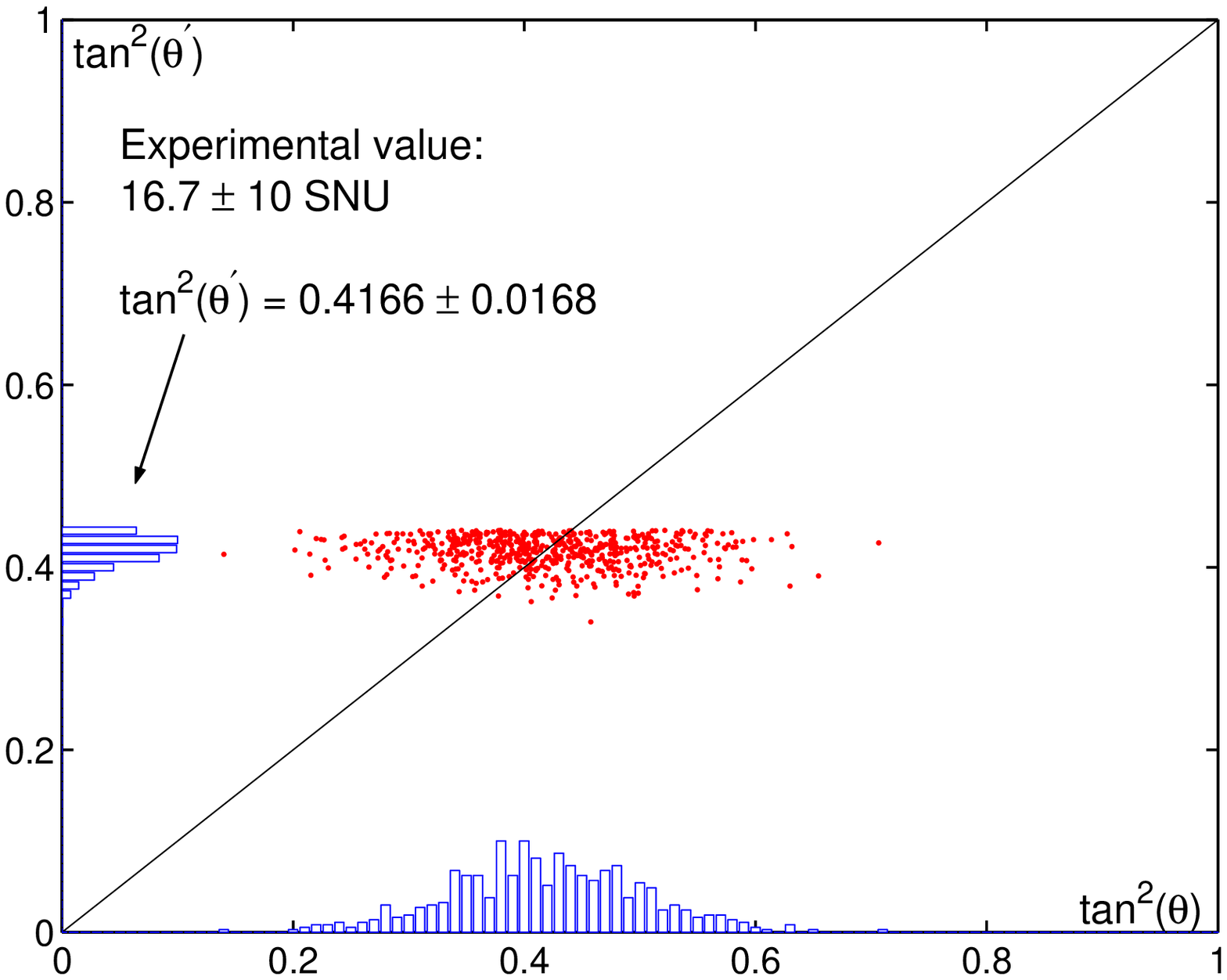}
\caption{The same as Fig.2 but for the experimental value $16.7
\pm 1\sigma$ SNU.}
\end{figure}

\begin{figure}[ht!]
\centering
\includegraphics[width=2.5in]{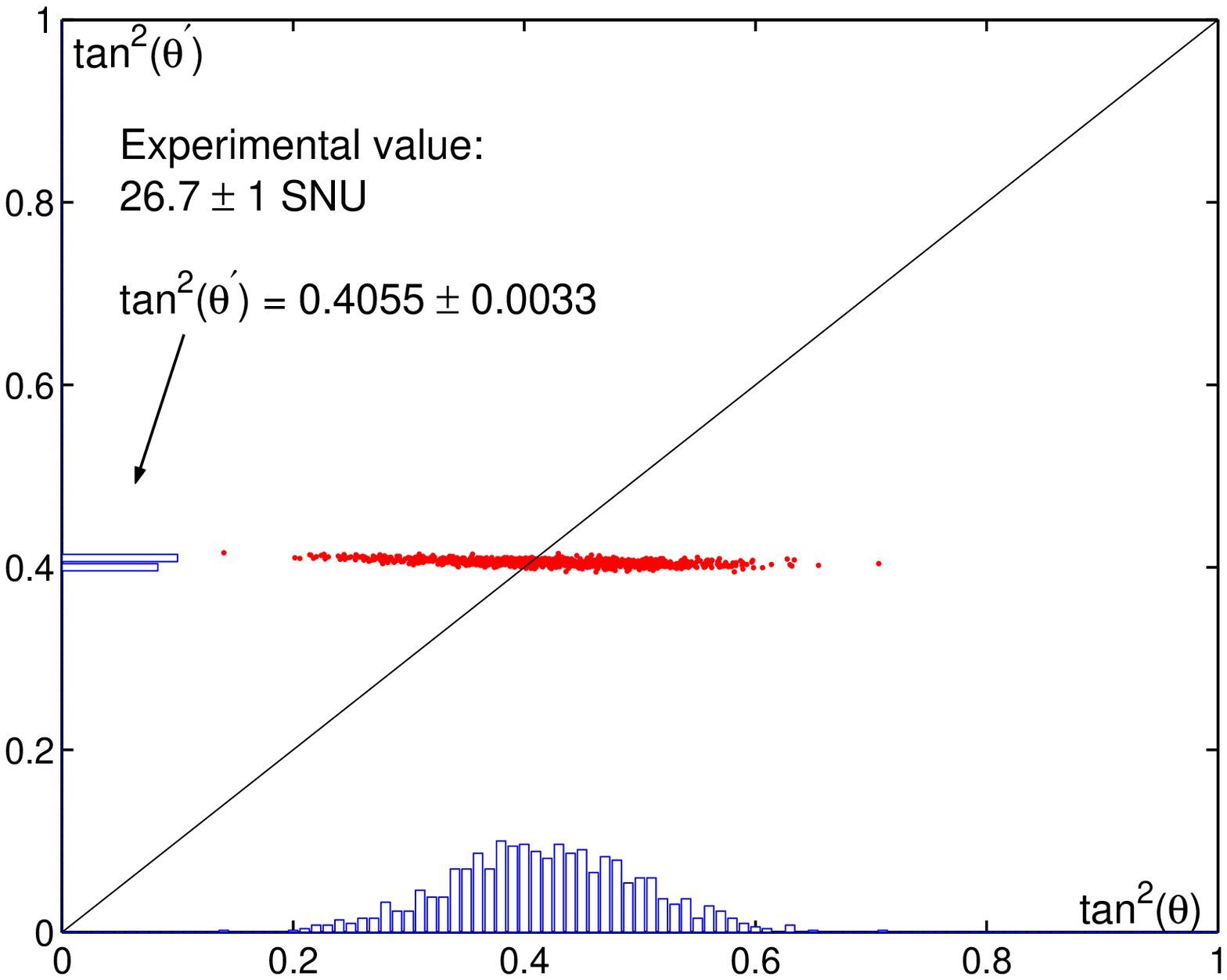}
\includegraphics[width=2.5in]{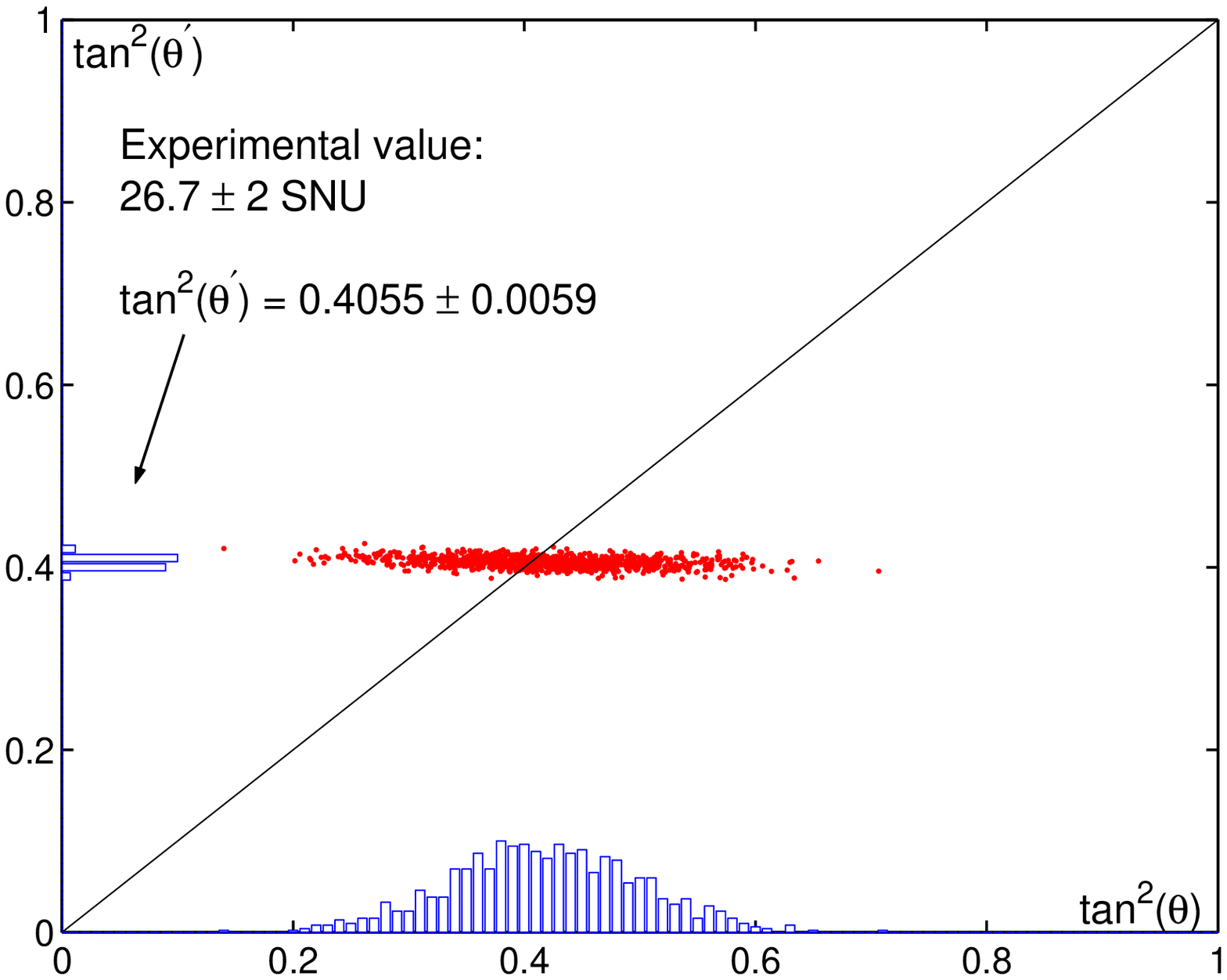}
\includegraphics[width=2.5in]{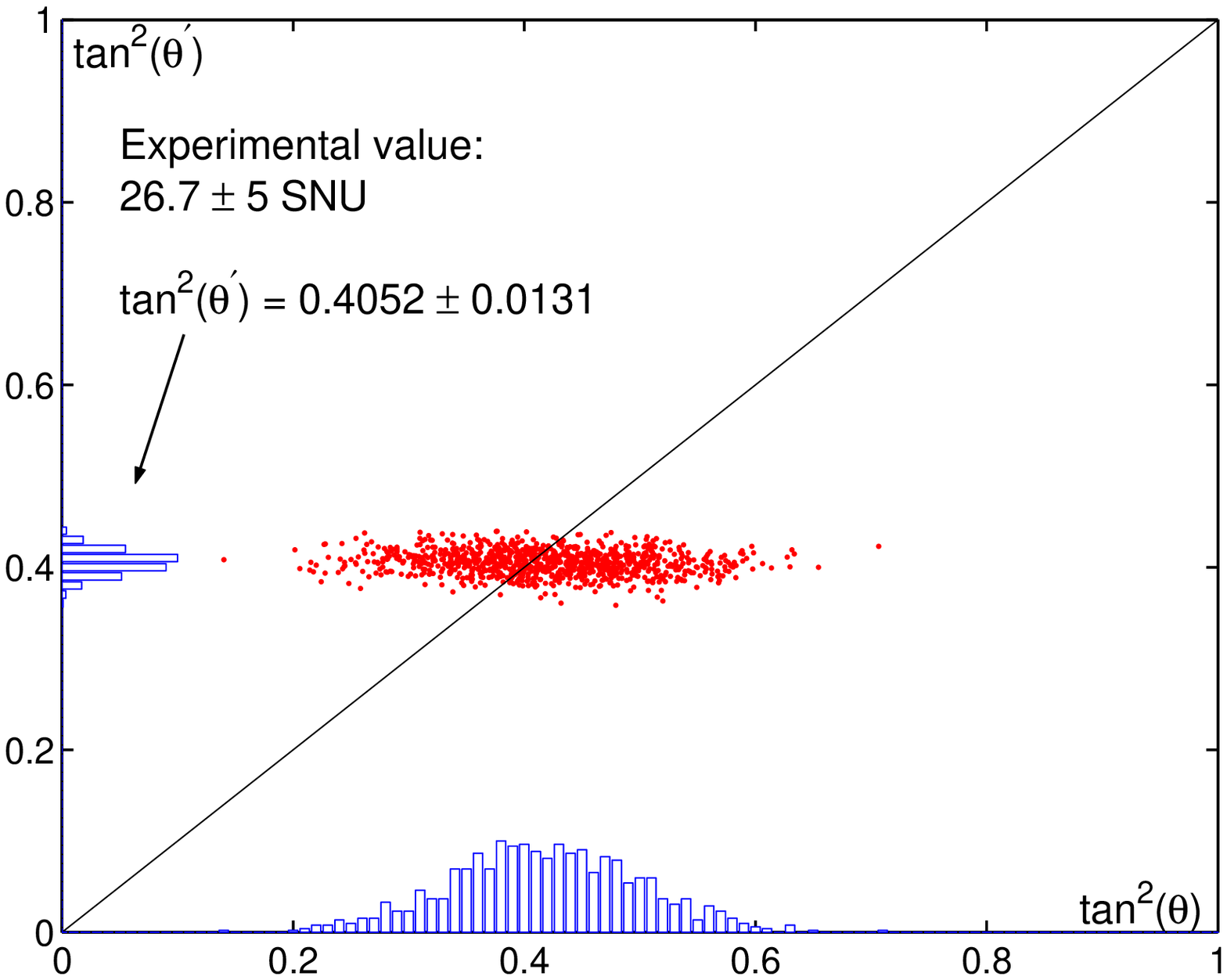}
\includegraphics[width=2.5in]{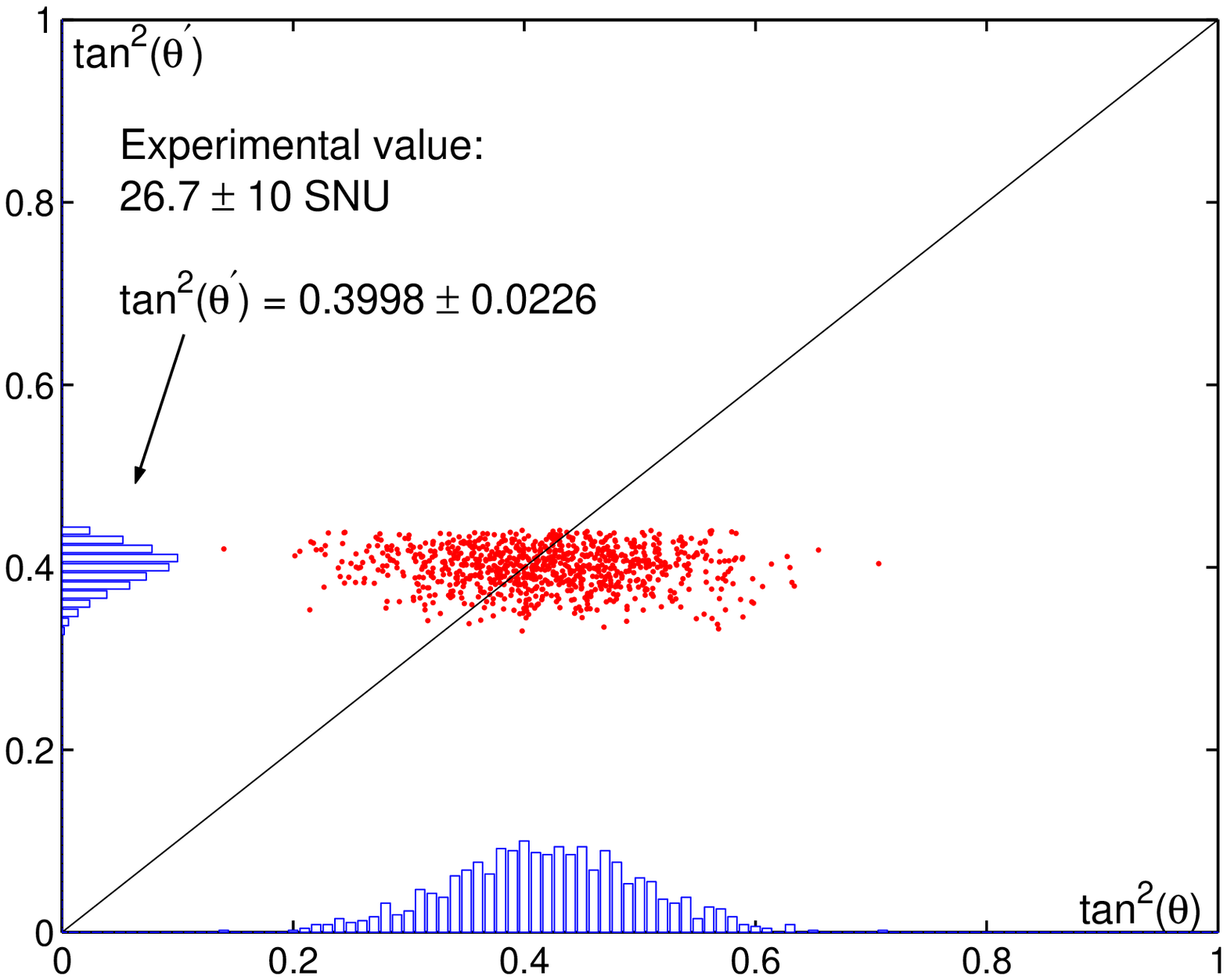}
\caption{The same as Fig.2 but for the experimental value $26.7
\pm 1\sigma$ SNU.}
\end{figure}

\begin{figure}[ht!]
\centering
\includegraphics[width=3in]{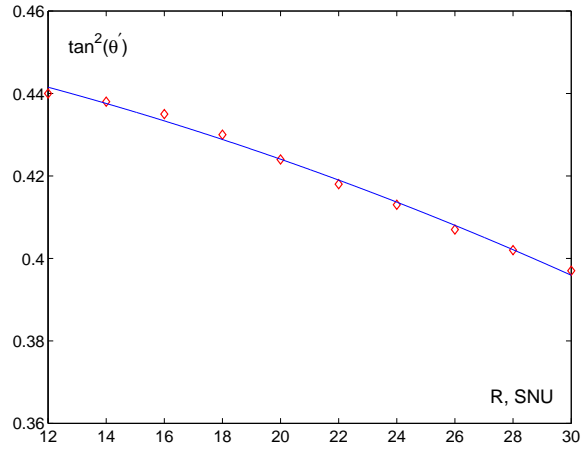}
\caption{The dependence of $tan^2\theta^{\prime}$ on the neutrino
capture rate measured in a lithium experiment.}
\end{figure}

\vskip 0.2in

\textbf{3. The conclusions important for a lithium experiment.}
\vskip 0.1in

For a future lithium experiment a very important point is that
even with the modest accuracy of about 5 SNU the result will be
very informative not only for establishing the role of CNO cycle
in the Sun but also for measuring a mixing angle. The accuracy of
about 5 SNU can be achieved with 10 tons of lithium using a
simplified counting system in a sense that it is aimed not to
counting each decay of $^7$Be what presumably is possible to do by
means of a cryogenic detector, but only a gamma line 0.478 KeV of
the excited state of daughter $^7$Li which is more convenient for
counting. Because the branching ratio of this line is only 10.4\%
the resulting efficiency of the counting by means of a low
background gamma-spectrometer will be very modest, of about 6\% .
But for the accuracy 5\% using 10 tons of metallic lithium it
still will be adequate to accomplish the task. If to take the
efficiency of extraction of beryllium from metallic lithium 80\% ,
the efficiency of counting 6\% , time of exposure 80 days and 4
Runs per year, the resulting accuracy will be about 5 SNU. Then 5
counts are expected in a Run which should be counted by a very low
background gamma-spectrometer within a time interval of about half
a year. The dangerous source of the background comes from the line
511 keV which is a well populated peak of the background spectra.
To discriminate this peak it is necessary to use a high resolution
detector with 4$\pi $ geometry, the best one for this aim is a
module composed of several high purity germanium detectors of the
kind planned to be used as a working module in a Majorana project.

This work was supported in part by the Russian Fund of Basic Research,
contract N01-02-16167-A and by the grant of Russia ``Leading Scientific
Schools'' LSS-1782.2003.2.

\end{document}